\documentstyle[aps,prbbib,twocolumn,epsf]{revtex}
\begin{document}
\medskip
\draft
\title{Optimal quantum pump in the presence of a superconducting
lead}

\author{Baigeng Wang $^1$ and Jian Wang$^{1,2,a}$}

\address{1. Department of Physics, The University of Hong Kong, 
Pokfulam Road, Hong Kong, China\\
2. Institute of Solid State Physics, Chinese Academy of Sciences,
Hefei, Anhui, China}
\maketitle

\begin{abstract}
We investigate the parametric pumping of a hybrid structure consisting
of a normal quantum dot, a normal lead and a superconducting lead. 
Using the time dependent scattering matrix theory, we have derived a 
general expression for the pumped electric current and heat current.
We have also derived the relationship among the 
instantaneous pumped heat current, electric current, and shot noise. 
This gives a lower bound for the pumped heat current in the hybrid 
system similar to that of the normal case obtained by Avron et al. 
%XX
%We have examined the heat current in the strong
%pumping regime and found that the double barrier structure can be
%functioned as an optimal pump in the presence of superconducting lead.
\end{abstract}


Since the seminar work of Brouwer,\cite{brouwer} the physics of parametric 
pumping has attracted increasing attention.\cite{switkes,zhou,wagner,avron,aleiner1,wei1,wang3,vavilov,brouwer2,kravtsov,aleiner,shutenko,levinson,buttiker1,wbg1,wang1}
Recently, Avron et al\cite{avron1} have considered the pumped heat current 
in the adiabatic regime and found a general lower bound for the heat
current. This defines an optimal pump if the heat current equals to the power 
of Joule heat dissipated during the pumping process\cite{avron1}. As a
consequence, the optimal pump is noiseless and charge transported is 
quantized. The physics of pumped heat current has also been investigated by
Moskalets and Buttiker\cite{buttiker} who have derived a general formula 
for the heat current in the weak pumping regime and the shot noise generated 
during the pumping process. In the strong pumping regime, the heat current
has been studied within the time-dependent scattering matrix 
theory\cite{wang5} and the existence of optimal pump has been examined. 
For chaotic quantum dots, Polianski et al\cite{vavilov1} have developed 
a time-dependent scattering matrix theory to account for the shot noise 
for parametric pumping and mesoscopic fluctuation for arbitrary temperature 
and beyond bilinear response. In this paper, we investigate the pumped
heat current for a normal superconducting (NS) hybrid system which consists 
of a normal quantum dot, a normal lead and a superconducting lead. In the
adiabatic regime, the energy of charged carriers (electron or hole) is
within the superconducting energy gap and hence physics of the Andreev 
reflection\cite{beenakker} dominates. We have derived a general expression 
for the pumped electric current and heat current in the presence of 
superconducting lead which is valid at finite pumping amplitude and 
finite temperature. Our theory is based on the time-dependent scattering 
matrix theory\cite{vavilov}. Since our theory is perturbative in nature, 
going beyond the adiabatic regime, one can in principle obtain the pumped 
electric current and heat current to any order in frequency. In the 
adiabatic regime, we have also derived a relationship among the 
instantaneous heat current, 
electric current, and the shot noise. This sets a lower bound for the 
pumped heat current. Similar to the normal system\cite{avron1}, a quantum
pump will be optimal if the pumped heat current reaches its lower bound. As
a result, the charge transported will be quantized and the system is 
noiseless just like the normal system. We have also compared with the 
pumped heat current of NS structure with that of normal structure. For 
a single pumping potential, the total heat currents generated are the same 
for NS and normal systems. For two pumping potentials the pumped heat 
current for NS system can be either larger or smaller than that of normal 
system depending on the phase difference between two pumping potentials. 

For the purpose of presentation, we consider the pumped electric current 
first. We start with the general definition for the electric current of 
type $\alpha$ (electron or hole) in the left normal lead in scattering 
matrix theory ($\hbar=q=1$),\cite{buttiker} 

\begin{equation}
I_{e,L\alpha} = \lim_{\Delta t -> \infty}
\frac{1}{\Delta t}\int^{\Delta t}_0 dt <{\hat I}_{e,L\alpha}>
\end{equation}
where $<...>$ denotes the quantum average and ${\hat I}_{e,L\alpha}$ 
is the electric current operator of type $\alpha$ in the left lead,
\begin{equation}
{\hat I}_{e,L\alpha} = q_\alpha [{\hat b}^{\dagger}_{L\alpha}(t) 
{\hat b}_{L\alpha}(t) - {\hat a}^\dagger_{L\alpha}(t) 
{\hat a}_{L\alpha}(t)]
\label{electric}
\end{equation}
Here the operators ${\hat b}_{L\alpha}$ and ${\hat a}_{L\alpha}$ are
annihilation operators for the outgoing and incoming carriers of type
$\alpha$ in the left lead and $q_\alpha=1,-1$ for $\alpha=e,h$. They 
are related by the scattering matrix,\cite{buttiker,datta}
\begin{equation}
{\hat b}_{L\alpha}(t) = \sum_\beta \int dt' {\cal S}_{\alpha \beta} (t,t')
{\hat a}_{L\beta}(t')
\label{bb}
\end{equation}
where the time-dependence of the scattering matrix ${\cal S}$ is 
due to the slowly time-varying pumping potential $X(t)$. 
The distribution function can be obtained by taking the quantum 
average,\cite{buttiker} 
\begin{equation}
<{\hat a}^\dagger_{L\alpha}(E) {\hat a}_{L\beta}(E')> = 
\delta_{\alpha \beta} \delta(E-E') f_L(E)
\label{fermi}
\end{equation}
where ${\hat a}_{L\alpha}(E)$ is the Fourier transform of 
${\hat a}_{L\alpha}(t)$ and $f_L(E)$ is the Fermi distribution 
function of the left lead. From Eqs.(\ref{electric}), (\ref{bb}), 
and (\ref{fermi}), the pumped electric current is given by

\begin{eqnarray}
I_{e,L\alpha} &=& \lim_{\Delta t -> \infty}
\frac{q_\alpha}{\Delta t}\int^{\Delta t}_0 dt \int dt_1
dt_2 \sum_\beta {\cal S}_{\alpha \beta}(t,t_1) \nonumber \\
&& f(t_1-t_2) {\cal S}^*_{\alpha \beta}(t,t_2) 
- q_\alpha\int \frac{dE}{2\pi} f(E)
\label{heat2}
\end{eqnarray}
where $f(t) \equiv \int (dE/2\pi) \exp(-iEt) f(E)$. 
After changing of the variable $t_0=(t_1+t_2)/2$ and $\tau=t_1-t_2$
and using the following Wigner transform for the scattering
matrix\cite{vavilov},

\begin{equation}
{\cal S}(t,t') = \int \frac{dE}{2\pi} e^{-iE(t-t')} 
{\cal S}(E,\frac{t+t'}{2})
\end{equation}
then Eq.(\ref{heat2}) becomes,

\begin{eqnarray}
&&I_{e,L\alpha} = \lim_{\Delta t -> \infty}
\frac{q_\alpha}{4\pi^2\Delta t}\int^{\Delta t}_0 dt \int dt_0
d\tau dE_1 dE_2 ~ f(\tau) \nonumber \\
&& e^{-iE_1(t-t_0-\tau/2)} e^{iE_2(t-t_0+\tau/2)} 
\sum_\beta {\cal S}_{\alpha \beta}(E_1, \frac{t+t_0}{2} 
+\frac{\tau}{4}) \nonumber \\
&& {\cal S}^*_{\alpha \beta}(E_2,\frac{t+t_0}{2}-\frac{\tau}{4})
-q_\alpha\int \frac{dE}{2\pi} f(E)
\end{eqnarray}
Changing the variables again to $\tau_1=t-t_0$ and $t'=(t+t_0)/2$ and
integrating over $\tau_1$, we obtain

\begin{eqnarray}
&&I_{e,L\alpha} = \lim_{\Delta t -> \infty}
\frac{q_\alpha}{2\pi\Delta t}\int^{\Delta t}_{-\Delta t} dt' 
d\tau dE ~ e^{iE \tau} f(\tau) \nonumber \\
&&\sum_\beta {\cal S}_{\alpha \beta}(E,t'+\tau/4) 
{\cal S}^*_{\alpha \beta}(E,t'-\tau/4) - q_\alpha\int \frac{dE}{2\pi} f(E) 
\label{i0}
\end{eqnarray}
Using the fact that 
\begin{eqnarray}
&&\lim_{\Delta t -> \infty}
\int^{\Delta t}_{-\Delta t} dt'
\sum_\beta {\cal S}_{\alpha \beta}(E,t'+\tau/4) 
{\cal S}^*_{\alpha \beta}(E,t'-\tau/4) \nonumber \\
&&=\lim_{\Delta t -> \infty}
\int^{\Delta t}_{-\Delta t} dt
\sum_\beta {\cal S}_{\alpha \beta}(E,t) e^{-(\tau/2)\partial_t} 
{\cal S}^*_{\alpha \beta}(E,t) 
\end{eqnarray}
Eq.(\ref{i0}) becomes
\begin{eqnarray}
&&I_{e,L\alpha} = \frac{q_\alpha}{\pi T_p}\int^{T_p}_{0} dt
\int dE ~ \{{\hat {\cal S}}(E,t) \nonumber \\
&&\times [f(E+i\partial_t/2)-f(E)] 
~ {\hat {\cal S}}^\dagger(E,t)\}_{\alpha \alpha}
\label{i1}
\end{eqnarray}
where $T_p$ is the period of the pumping cycle
and $\hat{{\cal S}}$ is a $2\times 2$ scattering matrix for NS structure
with matrix element ${\cal S}_{\alpha \beta}$ ($\alpha, \beta = e, h$). 
Eq.(\ref{i1}) is symbolic and is the central result of this paper. One
can in principle obtain the pumped electric current to any order in
frequency. For instance, to get the electric current up to $\omega$, 
it is enough to expand $f(E+i\partial_t/2)$ up to the first order in 
$\partial_t$, from which we obtain

\begin{eqnarray}
I_{e,L\alpha} &=& \frac{iq_\alpha}{2\pi T_p}\int^{T_p}_0 dt 
\int dE ~ \partial_E f \nonumber \\
&& [\partial_t {\hat {\cal S}}^\dagger(E,t)
{\hat {\cal S}}(E,t)]_{\alpha \alpha}
\label{final}
\end{eqnarray}
Note that from the unitary condition of the scattering matrix ${\hat
{\cal S}}$, we have 
\begin{equation}
\sum_\beta {\cal S}^*_{\alpha \beta} ~ {\cal S}_{\alpha \beta}=1
\label{uni}
\end{equation}
taking the derivative with respect to time, we obtain
\begin{equation}
\sum_\beta \partial_t {\cal S}^*_{\alpha \beta} ~ {\cal S}_{\alpha \beta}
+c.c.=0
\end{equation}
Hence ${\rm Im}[(\partial_t {\hat {\cal S}}^\dagger 
{\hat {\cal S}})_{\alpha \alpha}] = -i(\partial_t {\hat {\cal S}}^\dagger 
{\hat {\cal S}})_{\alpha \alpha}$. 
In the adiabatic regime, we have\cite{vavilov} $\partial_t 
{\cal S}_{\alpha \beta} = \sum_i[\partial_{X_i} {\cal S}_{\alpha \beta} 
\partial_t X_i + \partial_{{\dot X}_i} {\cal S}_{\alpha \beta} 
\partial_t {\dot X}_i + ...]$ where ${\dot X} \equiv dX/dt$. 
Up to the order $\omega$, we can neglect the contribution from 
$\partial_{{\dot X}_i} s_{\alpha \beta}$. 
At zero temperature, Eq.(\ref{final}) becomes,

\begin{equation}
I_{e,L\alpha} = \frac{iq_\alpha}{2\pi T_p}\int^{T_p}_0 dt 
[\partial_{X_i} {\hat {\cal S}} 
~ {\hat {\cal S}}^\dagger]_{\alpha \alpha} ~ \partial_t X_i
\label{final1}
\end{equation}
which agrees with the theory of nonequilibrium Green's 
function\cite{foot1}. 

Now we proceed to derive the pumped heat current for NS structure. 
We note that the heat current is defined as the particle current
multiplied by the energy measured from the Fermi level, we thus have
from Eq.(\ref{i1}), 
\begin{eqnarray}
&&I_{q,L\alpha} = \frac{1}{\pi T_p}\int^{T_p}_{0} dt
\int dE ~ (E-E_F) ~ \{{\hat {\cal S}}(E,t) \nonumber \\
&&\times [f(E+i\partial_t/2)-f(E)] 
~ {\hat {\cal S}}^\dagger(E,t)\}_{\alpha \alpha}
\label{i2}
\end{eqnarray}
Expanding the heat current up $\partial_t^2$ and after some algebra, we 
finally obtained the heat current up to $\omega^2$,

%\begin{eqnarray}
%I_{q,L\alpha} &=& \frac{1}{8\pi T_p}\int^{T_p}_0 dt 
%\sum_\beta \sum_{ij} \partial_{X_i} {\cal S}_{\alpha \beta}  \nonumber \\
%&&~ \partial_{X_j} {\cal S}^\dagger_{\alpha \beta} ~ \partial_t X_i
%~ \partial_t X_j
%\label{final2}
%\end{eqnarray}
\begin{equation}
I_{q,L\alpha} = \frac{-1}{8\pi T_p}\int^{T_p}_0 dt \int dE ~ \partial_E f
(\partial_t {\hat {\cal S}}
~ \partial_t {\hat {\cal S}}^\dagger)_{\alpha \alpha}
\label{final2}
\end{equation}
Now we derive the relationship between instantaneous electric current 
(denoted as $I_e(t)=-iq_\alpha (\partial_t {\hat {\cal S}}^\dagger 
{\hat {\cal S}})_{\alpha \alpha}$) and heat current ($I_q(t)=(\partial_t 
{\hat {\cal S}}^\dagger \partial_t {\hat {\cal S}})_{\alpha \alpha}$).
Now the instantaneous heat current becomes
\begin{eqnarray}
I_q(t)&&= (\partial_t {\hat {\cal S}}^\dagger ~ \partial_t {\hat 
{\cal S}})_{\alpha \alpha}
= (\partial_t {\hat {\cal S}}^\dagger ~ {\hat {\cal S}}  {\hat 
{\cal S}}^\dagger ~ \partial_t {\hat {\cal S}})_{\alpha \alpha} 
\nonumber \\
&&= \sum_\beta (\partial_t {\hat {\cal S}}^\dagger ~ {\hat 
{\cal S}})_{\alpha \beta} ({\hat {\cal S}}^\dagger ~ \partial_t 
{\hat {\cal S}})_{\beta \alpha}
\label{relation}
\end{eqnarray}
We see that the diagonal term in Eq.(\ref{relation}) is just the 
$I_e^2(t)$ and the off diagonal term is the shot noise $S_0(t)$ generated 
during the pumping process\cite{buttiker,avron1,foot4}. Therefore, we have 
the following relationship
\begin{equation}
I_q(t)=I_e^2(t)+S_0(t)
\label{opt}  
\end{equation}
or general lower bound for the heat current 
\begin{equation}
I_q(t) \geq I_e^2(t)
\end{equation}
The condition of optimal pump for NS structure is defined as $S_0(t)=0$.
Following Avron et al\cite{avron1} it is straightforward to show that 
the charge transported through the system per cycle is quantized 
if the quantum pump is optimal. 

Now we consider the weak pumping limit for a symmetric double barrier 
structure in the presence of superconducting lead. The double barrier 
structure is modeled by potential $V(x) = X_1(t) \delta(x+a) + X_2(t) 
\delta(x-a)$ where $X_1(t) = X_0+X_1 \sin(\omega t)$ and 
$X_2(t) = X_0+ X_2 \sin(\omega t +\phi)$. In this limit, it is easy to 
show that the pumped electric current and heat current are given,
respectively, by
\begin{equation}
I_{e,L\alpha} = \frac{\omega q_\alpha \sin\phi X_1 X_2}{2\pi} 
{\rm Im}[(\partial_{X_1} {\hat {\cal S}}^\dagger
\partial_{X_2} {\hat {\cal S}})_{\alpha \alpha}]
\label{weak}
\end{equation}
and 
\begin{eqnarray}
I_{q,L\alpha} &=& \frac{\omega^2}{16\pi} [
X_1^2 \partial_{X_1} {\hat {\cal S}}^\dagger
\partial_{X_1} {\hat {\cal S}}
+X_2^2 \partial_{X_2} {\hat {\cal S}}^\dagger
\partial_{X_2} {\hat {\cal S}}
\nonumber \\
&+& 2\cos\phi ~ X_1 X_2 {\rm Re} (\partial_{X_1} 
{\hat {\cal S}} \partial_{X_2} {\hat {\cal S}}^\dagger)]_{\alpha \alpha}
\label{weak1}
\end{eqnarray}
In Eqs.(\ref{weak}) and (\ref{weak1}), we have set $X_i=0$ in ${\cal S}$
after the partial derivatives. Now we will calculate the pumped heat
current $I_{q,L\alpha=e}$ for the double barrier NS system.
For the NS system, the scattering matrix ${\cal S}_{ee}$ and 
${\cal S}_{he}$ are given by\cite{beenakker,lesovik}
\begin{equation}
\hat{{\cal S}} = \hat{S}_{11} + \hat{S}_{12} (1 - \hat{R}_I
\hat{S}_{22})^{-1} \hat{R}_I \hat{S}_{21}
\label{lesovik}
\end{equation}
where
\begin{equation}
\hat{S}_{ij}(E)=
\left(
\begin{tabular}{cc}
$S_{ij}(E)$ & $0$ \\ 
$0$ & $S_{ij}(-E)$
\end{tabular}
\right)
\end{equation}
with $S_{ij}$ being usual scattering matrix for the normal structure. 
$\hat{R}_I=\alpha \sigma_x$ is the $2\times 2$ 
scattering matrix at NS interface with off diagonal matrix element 
$\alpha$. Here $\alpha = (E-i\nu \sqrt{\Delta^2-E^2})/\Delta$ with 
$\nu=1$ when $E>-\Delta$ and $\nu=-1$ when $E<-\Delta$. In
Eq.(\ref{lesovik}), the energy $E$ is measured relative to the chemical
potential $\mu$ of the superconducting lead.  Eq.(\ref{lesovik}) 
has clear physical meaning\cite{lesovik}. The first term is the
direct reflection from the normal scattering structure and the second
term can be expanded as $\hat{S}_{12} \hat{R}_I \hat{S}_{21} + \hat{S}_{12} 
\hat{R}_I \hat{S}_{22} \hat{R}_I \hat{S}_{21} + ...$ which is
clearly the multiple Andreev reflection in the hybrid structure.
From Eq.(\ref{lesovik}) we obtain the well known expressions for the
scattering matrix ${\cal S}_{ee}$ and ${\cal S}_{he}$\cite{beenakker}
\begin{equation}
{\cal S}_{ee}(E) = S_{11}(E) + \alpha^2 S_{12}(E) S_{22}^*(-E) M_e 
S_{21}(E)
\label{see}
\end{equation}
and
\begin{equation}
{\cal S}_{he}(E) = \alpha S_{12}^*(-E) M_e S_{21}(E)
\label{she}
\end{equation}
with $M_e = [1- \alpha^2 S_{22}(E) S_{22}^*(-E)]^{-1}$.
In the case of parametric pumping,  we assume that the Fermi energy 
is in line with the chemical potential of superconducting lead, so $E=0$ 
and $\alpha =-i$. For the symmetric NS system at resonance, we have 
$S_{11}=0$ and $S_{12}= e^{-2ika}$ in the absence of pumping potential. 
Therefore, from Eqs.(\ref{see}) and (\ref{she}), we have 
\begin{equation}
\partial_{X_{1/2}} {\cal S}_{ee} = \partial_{X_{1/2}} 
S_{11} - S_{12}^2 \partial_{X_{2/1}} S_{11}^*
\label{d1}
\end{equation}
and
\begin{equation}
\partial_X {\cal S}_{he} = -i (\partial_X S^*_{12} S_{12}+c.c.)
\label{d2}
\end{equation}
where we have used the fact that $\partial_{X_1} S_{22} = 
\partial_{X_2} S_{11}$. Using Fisher-Lee relation\cite{lee} 
$S_{\alpha \beta} = - \delta_{\alpha \beta} + i v
G^r_{\alpha \beta}$ and the Dyson equation $\partial_{X_j} 
G^r_{\alpha \beta} = G^r_{\alpha j} G^r_{j \beta}$\cite{gasparian}, we have 
$\partial_{X_1} S_{11} =ivG^r_{11} G^r_{11}=-i/v$, $\partial_{X_2}
S_{11} = iv G^r_{12} G^r_{21} = -iS_{12}^2/v$, and $\partial_{X_{1/2}}
S_{12} = -iS_{12}/v$ with the velocity $v=2k$. Thus from Eqs.(\ref{d1}) and
(\ref{d2}, we have $\partial_{X_j} {\cal S}_{ee} = 2 \partial_{X_j} S_{11}$ 
and $\partial_X {\cal S}_{he}=0$. From Eq.(\ref{weak1}), we obtain
\begin{equation}
I^{NS}_{q,L e} = \frac{\omega^2}{16k^2}[X_1^2+X_2^2+2\cos\phi ~ X_1 X_2 
\cos4ka]
\label{ns1}
\end{equation}
which should be compared with the pumped heat current in the normal case
\begin{equation}
I^{N}_{q,L} =I^N_{q,R}= \frac{\omega^2}{32k^2}[X_1^2+X_2^2+\cos\phi ~ X_1
X_2 (1+\cos4ka)]
\label{normal1}
\end{equation}
We note that in the NS system, the heat current flows out only through
the normal lead while for normal system, the heat current pumps out
through both leads. Comparing Eqs.(\ref{ns1}) and (\ref{normal1}), we have

\begin{eqnarray}
I^N_{q,L}+I^N_{q,R}-I^{NS}_{q,Le}&=& \frac{\omega^2}{16k^2} \cos\phi ~ X_1 
\nonumber \\
&\times& X_2 ~ (1-\cos4ka)
\end{eqnarray}
Hence the total pumped heat current generated in the normal system is can be 
either larger or smaller than that in the NS system depending on the phase 
difference of two pumping potentials. For a single pump, by setting $X_2=0$ 
in Eqs.(\ref{ns1}) and (\ref{normal1}), we see that the total pumped heat 
currents are the same for both NS and normal systems. This is different from
the pumped electric current where in the weak pumping regime at resonance,
the electric current for NS system is four time larger than that of normal
system\cite{wang3}.

In summary, we have derived a general expression for the pumped electric
current and heat current in the presence of superconducting lead using the 
time-dependent scattering matrix theory. Our theory is valid at finite 
pumping amplitude and can be applied to the multi-channel systems. 
Since our theory is perturbative in nature, we can expanding Eq.(\ref{i2}) 
to the higher order in frequency and hence approach to the nonadiabatic 
regime. Our theory can also be easily extended to the case of multi-terminal 
structures. Although our expression is derived for NS system, it is also
valid the normal system as well by simply replacing the NS scattering
matrix ${\cal S}_{\alpha \beta}$ with $\alpha \beta=e,h$ by normal
system scattering matrix $S_{ij}$ with $i,j=1,2$ in Eqs.(\ref{i1}) and
(\ref{i2}). For the NS system, we have found the lower bound for the 
pumped heat current similar to that of Avron et al\cite{avron1} for the
normal system. As a result, the optimal pump can exist for NS system as 
well. In the weak pumping limit, we have examined the pumped heat current 
for NS system for a double barrier structure at resonance. 
For two parameter pump, we found that the total pumped heat current for NS 
structure can be larger or smaller than that of normal structure depending 
on the phase different between two pumping parameters.

\section*{Acknowledgments}
We gratefully acknowledge support by a RGC grant from the SAR Government 
of Hong Kong under grant number HKU 7091/01P and a CRCG grant from The
University of Hong Kong.

\medskip
\bigskip
\bigskip
\noindent{$^{a)}$ Electronic mail: jianwang@hkusub.hku.hk}

%\begin{figure}
%\caption{
%%%% note that in caption, \ref is not allowed.
%The transmission coefficient (solid line) and heat current as a
%function of Fermi energy at different phase differences between two
%}
%\end{figure}


\begin{thebibliography}{00} 

\bibitem{brouwer}
P.W. Brouwer, Phys. Rev. B {\bf 58}, R10135 (1998). 

\bibitem{switkes}
M. Switkes, C. Marcus, K. Capman, and A.C. Gossard, Science {\bf 283}, 
1905 (1999). 

\bibitem{zhou}
F. Zhou, B. Spivak, and B.L. Altshuler, Phys. Rev. Lett. {\bf 82}, 
608 (1999).

\bibitem{wagner}
M. Wagner, Phys. Rev. Lett. {\bf 85}, 174 (2000).

\bibitem{avron}
J.E. Avron, A. Elgart, G.M. Graf, and L. Sadun, Phys. Rev. B {\bf
62}, R10618 (2000).

\bibitem{aleiner1}
I.L. Aleiner, B.L. Altshuler, and A. Kamenev, Phys. Rev. B {\bf 62},
10373 (2000). 

\bibitem{wei1}
%Y.D. Wei, J. Wang, and H. Guo, Phys. Rev. B {\bf 62}, 9947 (2000);
%Y.D. Wei, J. Wang, H. Guo, and C. Roland, Phys. Rev. B {\bf 64},
%115321 (2001);
%Y.D. Wei and J. Wang, cond-mat/0207473.
Y.D. Wei et al, Phys. Rev. B {\bf 62}, 9947 (2000);
Phys. Rev. B {\bf 64}, 115321 (2001); cond-mat/0207473.

\bibitem{wang3}
%J. Wang, Y.D. Wei, B.G. Wang, and H. Guo, Appl. Phys. Lett. {\bf
%79}, 3977 (2001). 
J. Wang et al, Appl. Phys. Lett. {\bf 79}, 3977 (2001). 

\bibitem{vavilov}
%M.G. Vavilov, V. Ambegaokar, and I.L. Aleiner, Phys. Rev. B {\bf
%63}, 195313 (2001). 
M.G. Vavilov et al, Phys. Rev. B {\bf 63}, 195313 (2001). 

\bibitem{brouwer2}
P.W. Brouwer, Phys. Rev. B {\bf 63}, 121303 (2001); 
M.L. Polianski and P.W. Brouwer, Phys. Rev. B {\bf 64}, 075304
(2001).

\bibitem{kravtsov}
X.B. Wang and V.E. Kravtsov, Phys. Rev. B {\bf 64}, 033313 (2001).

\bibitem{aleiner}
I.L. Aleiner and A.V. Andreev, Phys. Rev. Lett. {\bf 81}, 1286 (1998). 

\bibitem{shutenko}
T.A. Shutenko, I.L. Aleiner, and B.L. Altshuler, Phys. Rev. B {\bf
61}, 10366 (2000). 

\bibitem{levinson}
Y. Levinson, O. Entin-Wohlman, and P. Wolfle, Physica A 
{\bf 302}, 335 (2001). 

\bibitem{buttiker1}
M. Moskalets and M. Buttiker, Phys. Rev. B {\bf 64}, 201305 (2001).

\bibitem{wbg1}
B.G. Wang, J. Wang, and H. Guo, Phys. Rev. B {\bf 65}, 073306 (2002). 

\bibitem{wang1}
J. Wang and B.G. Wang, Phys. Rev. B {\bf 65}, 153311 (2002); 
B.G. Wang and J. Wang, Phys. Rev. B {\bf 65}, 233315 (2002); 
%J.L. Wu, B.G. Wang, and J. Wang, cond-mat/0204570.
J.L. Wu et al, cond-mat/0204570.

\bibitem{avron1}
J.E. Avron, A. Elgart, G.M. Graf, and L. Sadun, Phys. Rev. Lett. 
{\bf 87}, 236601 (2001).

\bibitem{buttiker}
M. Moskalets and M. Buttiker, cond-mat/0201259.

\bibitem{wang5}
B.G. Wang and J. Wang, cond-mat/0204067.

\bibitem{vavilov1}
M.L. Polianski, M.G. Vavilov, and P.W. Brouwer, Phys. Rev. B {\bf 65},
245314 (2002).

\bibitem{beenakker}
C.W.J. Beenakker, Rev. Mod. Phys. {\bf 69}, 731 (1997).

\bibitem{datta}
M.P. Anantram and S. Datta, Phys. Rev. B {\bf 53}, 16390 (1996).

\bibitem{foot1}
There is a typo mistake in Eq.(2) of Ref.\onlinecite{wang3} that the
minus sign should be the plus sign. 

\bibitem{foot4}
When $\hbar \omega << k_B T$, the shot noise is $S(t) = S_0(t)/(4\pi k_B T)
= |\partial_t {\cal S}^*_{he} {\cal S}_{he}|^2/(4\pi k_B T)$. 
 
\bibitem{lesovik}
G.B. Lesovik, A.L. Fauchere, and G. Blatter, Phys. Rev. B {\bf 55},
3146 (1997). 

\bibitem{lee}
D. S. Fisher and P. A. Lee, Phys. Rev. B {\bf 23}, 6851 (1981). 

\bibitem{gasparian}
V. Gasparian, T. Christen, and M. Buttiker, Phys. Rev. A {\bf 54}, 4022 
(1996). 

%\bibitem{yip}
%M.K. Yip, J. Wang, and H. Guo, Z. Phys. B: Condens. Matter 
%{\bf 104}, 463 (1997). 

%\bibitem{foot2}
%For the GaAs system with $a=1000A$, the energy uint is $E=56 \mu eV$. 

%\bibitem{foot3}
%The power of Joule heat is defined as $(1\pi/2T_p) \int_0^{T_p} dt 
%(dQ/dt)^2$ where $Q = \int_0^{T_p} dt (dQ/dt)$ gives the pumped charge
%and $I_p = Q/T_p$ the pumped electric current.

\end{thebibliography}
\end{document}